\documentclass[11pt]{article}
\usepackage[english]{babel}
\usepackage[dvips]{graphicx }
\usepackage{cite}
\newcommand{\be}{\begin{equation}}
\newcommand{\ee}{\end{equation}}
\newcommand{\bea}{\begin{eqnarray}}
\newcommand{\eea}{\end{eqnarray}}
\def\vh{\varphi}

\title{ Astrophysics in relative units as the theory of a conformal brane}

\author{V.N. Pervushin, D.V. Proskurin, V.A. Zinchuk, A.G. Zorin\\
Bogoliubov Laboratory of Theoretical Physics,\\
Joint Institute for Nuclear Research, 141980 Dubna, Russia}

\begin{document}

\maketitle

%\date{\today}

\begin{abstract}
 The latest astrophysical data on the Supernova
 luminosity-distance -- redshift relations, primordial nucleosynthesis, value of
 Cosmic Microwave Background-temperature, and baryon asymmetry  are considered as
 an evidence of  relative measurement standard, field nature of time,
  and conformal symmetry  of the physical world.
  We show how  these principles of description of the universe
   help   modern quantum field theory to explain the creation
   of the universe, time, and matter from the physical vacuum as a state with the lowest energy.
\end{abstract}

%\tableofcontents

\section{Introduction}

  Observations and measurements of the physical parameters of cosmic objects
  and their theoretical
 interpretation within the framework of  modern physical theories
 of space, time, and matter
 (general relativity \cite{f22} and
 Standard Model of elementary particles \cite{linde})
 enable one to describe the history of cosmic evolution of the universe in
 the whole \cite{Chernin}.

 In particular,   observational data on  the dependence of the
  redshifts  $z$ of spectral lines of atoms on cosmic objects
 from their distance up to the Earth \cite{hubble}, and the new data
 \cite{snov,sn1997ff} for large values of  redshifts
 $z\sim 1,~z=1,7$ testify to that our  universe is mainly filled with
  not a massive ``dust'' of far and,
 therefore, invisible Galaxies, but with mysterious substance of a much different
 nature, with a different equation of state called {\it
 Quintessence} \cite{Q}. The data including primordial nucleosynthesis
  and the chemical evolution
of the matter in the universe (described in the nice book by Weinberg
\cite{three}) point to a
definite equation of state of the matter in the universe.
This equation helps us to determine a kind of matter taking part in the cosmic
evolution of the redshift.
The data on the visible number of
particles (baryons, photons, neutrinos, etc.) testify to that
the visible baryon matter gives only  $0.03$ part
 of the critical density $\rho_{\rm cr}$ of
the observational cosmic evolution  \cite{fuk}.
The data on the Cosmic Microwave Background radiation with
the temperature $2.7 K$ and its fluctuations \cite{33} give information
about the evolution of the early universe.

 Beginning with the pioneer papers by Friedmann \cite{f22} and ending with the
  last papers on inflationary model \cite{linde}
 of the Hot Universe Scenario \cite{gamow},
 all observational
  data are interpreted in the theoretical cosmology as some evidence
 of the expanding universe. Here, it is necessary  to clearly distinguish
 the expansion  of theoretical intervals from the  expansion of
 ``measurable intervals''  obtained by
  matching with a particular measurement standard.

 Not all clearly understand that this
  treatment of the Friedmann interval as a measurable one
   is true, if there are ``absolute'' units
  that do not expand together with the cosmic scale factor $a(t)$
  in the  universe, because an observer can measure only a ratio of
  any physical quantities and the units.
 Such a conjecture about
 the absolute measurement standard, irrespective of  how it will
 be selected (as one of the 40.000.000th part of Parisian
 Meridian or as a definite number of wave lengths
 of a spectral line of an isotope of krypton--86
  \cite{jay}), contradicts the  Einstein cosmological principle, according
   to which  ``any of the averaged characteristics of
 space environment does not select preferential position or
 preferential directions in the space'' \cite{1917}.

 The real situation is even more complicated.
 Not all clearly realize  that modern
 cosmology  in fact  utillizes the dual standard in
 describing the phenomenon of cosmic
  evolution of  photons emitted by  a massive matter on a far cosmic object.

 As soon as the cosmic photon has been carved out from atom, there
 are two distance scales:
 the wave length of a photon and the size of an atom that is determined  by its  mass.
  The observer can measure only evolution of a dimensionless ratio of
 the size of
 the atom, emitting a photon on a far cosmic object, to the wave length of
 this photon. These measurements irrefutably testify only to
 a permanent magnification of this dimensionless ratio.
 However,  these measurements cannot tell us
 what exactly is augmented: the wave length of a photon, or the mass of an atom
  emitting this photon. Thus, the observer who selected
 the absolute measurement standard of length states
  that the wave length is augmented;
 and who selected the relative one,  that the mass is augmented.

 The relative units  are used  in observational cosmology
 to determine initial data for cosmic photons  flying to an observer \cite{039},
 whereas the absolute units  are utilized for interpretation of
 observational data in theoretical cosmology.

 Theoretical cosmology considers the description in the
 relative units only as a mathematical method of solving problems,
 underlining that there are two mathematically
 equivalent  versions of the theoretical description of cosmological
 data in the form of two mathematically equivalent versions of
 general relativity and Standard Model. By virtue of this
 equivalence the usage of the dual standard in  cosmology does not
 lead  to conflicts and  enables one to reformulate the theory by
 treating the relative quantities as measurable ones, and the absolute ones as
 a mathematical tool  of solving problems.
 As a result we can recalculate all astrophysical data in
 the relative units, including the conformal time, coordinate
 distance, and constant temperature $T_c$, so that the z-history of temperature
 becomes the z-history of masses.

 The attempt of this recalculation  made in \cite{039,plb,039a,ppgc,114,02,295}
   has  shown, that the symmetry of equations of motion
   of the theory in the relative units
  increases, and number of phenomenological parameters decreases.
  This choice of the relative units results in a number of
 coincidences of parameters of cosmic evolution and elementary particle physics
 which  could be considered
 as random ones if such coincidences were not so large.

The purpose of the present paper is the description of the results and
 consequences of the {\it relative measurement standards}
  which   expand  together with the universe.

\section{Astrophysical data in the relative units}

 Theoretical cosmology is based on general relativity and the Standard
 Model of elementary particles constructed similarly to the Faraday - Maxwell
 electrodynamics.

  Maxwell revealed that the description of results
   of experimental measurement of  electromagnetic phenomena
   by the field theory equations
 depend on the definition of measurable quantities in the theory
  and the choice of their measurement standard.
 In the introduction
  to his {\it A Treatise on Electricity and Magnetism} \cite{Maxwell}
  Maxwell wrote:
{\it ``The most important aspect
 of any phenomenon from  mathematical
 point of view  is that of a measurable quantity.
  I shall therefore consider electrical phenomena
  chiefly with a  view to their measurement,
 describing the methods of measurement, and
 defining the  standards on
  which they depend''.}

Defining a measurable interval of the length as the ratio
\be\label{gr3}
{ ds}^2_{\mbox{\tiny MEASUREMENT}}=
\frac{{ds}^2_{\mbox{\tiny THEORY}}}{{[\mbox{\tiny STANDARD}]^2}},
\ee
 one need to point out its measurement standard.
  In  modern physics such a measurement standard of  length is the Parisian
 meter equal to a particular number of lengths of a light wave of a concrete
 spectral line of the  krypton isotope - 86 \cite{jay}.

 Physical cosmology based on the expanding interval \cite{f22}
\be\label{gr1}
{ds}^2_{\mbox{\tiny  THEORY}}={ (dt)^2-a^2(t)
\left[(dx^1)^2+(dx^2)^2+(dx^3)^2\right]},
\ee
with the scale factor $a(t)$ uses two standards: the relative and absolute.

Observational conformal cosmology  (CC) uses the relative Parisian meter
\be\label{rpm}
{\rm Relative~Parisian~Meter} =1 {\rm m}\times a(t)
\ee
for  measurements of {\it all lengths}
 with the corresponding conformal interval of the space-time
\be\label{gr3c} { ds}^2_{\mbox{\tiny
MEASUREMENT}}={ds}^2_{\mbox{\tiny
THEORY}}/a^2(t)=(d\eta)^2-(dx^i)^2 \ee of the cosmic photons
flying on the light cone to an observer. This interval is given in
terms of the conformal time $d\eta=dt/a(t)$ and coordinate
distance.

While in the theoretical standard cosmology  (SC) one proposes
that all lengths in the universe are measured with respect to the
absolute Parisian meter \be\label{apm} {\rm
Absolute~Parisian~Meter} =1 {\rm m}. \ee In the standard cosmology
the cosmic factor scales all distances {\it besides
  the Parisian meter}~(\ref{apm}).
  Nobody can explain why the measurement standard is so distinguished
  in the expanding universe.

%In the standard cosmology, an absolute measured distance  is
%defined as the product of the scale factor and the coordinate
%distance $X^i=a(t)x^i$. This product can be treated as a conformal
%transformation that leads to the theory  with constant mass. This
%theory is mathematically equivalent to the theory with a relative
%interval and variable mass as all solutions of the second theory
%can be received from those of the first theory by the conformal
%transformation. However, the mathematical equivalence does not
%mean physical equivalence of the absolute and relative units. When
%we assert that the cosmic factor was equal to the square root of
%``time'' in the epoch of the primordial nucleosynthesis, the
%question appears: what time does an observer measure by his watch,
%and what time is identified with the time of chemical evolution?

 Using the light cone equation $d t =  a(t)d r_c$
  one can find the coordinate distance - conformal time
  relation
  \be \label{dse4}
 r_c(\eta)=\sqrt{(x^1)^2 + (x^2)^2 + (x^3)^2}=\int\limits_{t}^{t_0}\frac{dt'}{a(t')}\equiv\eta_0-\eta,
 \ee
 where $\eta_0$ is the present-day value of the conformal time when a value of
 the scale factor is equal to unit $a(t)\Big{|}_{\eta=\eta_0}=1$.
 Therefore, the current cosmological time $\eta$ of a photon emitted by an atom
 at the coordinate distance $r_c$ is equal to the difference
 \be\label{1dse4}
 \eta=\eta_0-r_c.
 \ee
 Observational data testifies  that the energy of
  cosmic photons $E(\eta)$ depends on the coordinate distance~(\ref{dse4}).
 The  energy  of the cosmic photons $E(\eta_0-r_c)$ (emitted at the
 conformal time $\eta=\eta_0-r_c$) is always less then
 the similar energy of the Earth photons $E_0=E(\eta_0)$
 (emitted at the conformal time $\eta_0$):
 \be\label{Sch2}
 E(\eta)=\frac{E_0}{z(r_c)+1},
 \ee
  where
 \be\label{redshift}
 z(r_c) +1=\frac{E_0}{E(\eta_0-r_c)}\geq 1
 \ee
 is the redshift of
 the spectral lines $E$ of atoms at objects
 at the  coordinate distance $r_c$ in comparison with the
 the present-day spectral lines $E_0$ of atoms at the Earth.

In terms of the relative units (\ref{rpm}) and the conformal interval (\ref{gr3c})
 we reveal that
the measurable spatial volume of the universe is a constant
$V_{\rm {(r)}}$, while all masses including the Planck mass are
scaled by the cosmic scale factor \be \label{mass}
m_{(r)}(\eta)=m_0a(t)=m_0\tilde{a}(\eta). \ee

In this case, we get the  relation (\ref{Sch2}) as an eigenvalue
of solutions of the Schr\"odinger equation with the running mass
(\ref{mass}) \be\label{1Sch1} \left[\frac{\hat p^2}{2m_0
 \tilde{a}(\eta)}-\left(\frac{\alpha}{r}+E_{k}(\eta) \right) \right]\Psi_A=0.
 \ee
It is easy to check that in the limit of $E_k(\eta)\gg
\tilde{a}'/\tilde{a}=H(\eta)$ eigenvalue of the solution of this
equation is expressed throw the one of
 solution $E^0_{k}$ of a similar Schr\"odinger equation
 with constant masses $m_0$
 at $\tilde{a}(\eta_0)=1$
 \be\label{1Sch2}
 E_k(\eta)=\tilde{a}(\eta)E^0_{k}\equiv\frac{E^0_{k}}{z(r_c)+1},
 \ee
  where $E^0_{k}=-{m_0\alpha^2}/{k^2}$ is the spectrum of a Coulomb atom\footnote{The spectrum of photons emitted by atoms from
distant stars billion years ago remains unchanged during the
propagation and is determined by the mass of the constituents at
the moment of emission. When this spectrum is compared with the
spectrum of similar atoms on the Earth which, at the present time,
have larger masses, then a redshift is obtained.}.

It was shown~\cite{plb,039,ppgc} that the relative units give a
completely different physical picture of the evolution of the
universe than the absolute units of the standard cosmology. The
temperature history of the expanding universe
 copied in the relative units looks like the
history of evolution of masses of elementary particles in the cold universe
with a constant temperature of the cosmic microwave background.

%Theoretical standard cosmology  (SC) proposes that all lengths in
%the universe are measured with respect to the absolute Parisian meter
%\be\label{apm}
%{\rm Absolute~Parisian~Meter} =1 {\rm m}.
%\ee
%In the standard cosmology the cosmic factor scales all distances {\it besides
%  the Parisian meter}~(\ref{apm}).
%  Nobody can explain why the measurement standard is so distinguished
%  in the expanding universe.
%

\begin{figure}[ht]
\vspace{1cm}
 \begin{center}
 \includegraphics[width=0.65\textwidth,clip]{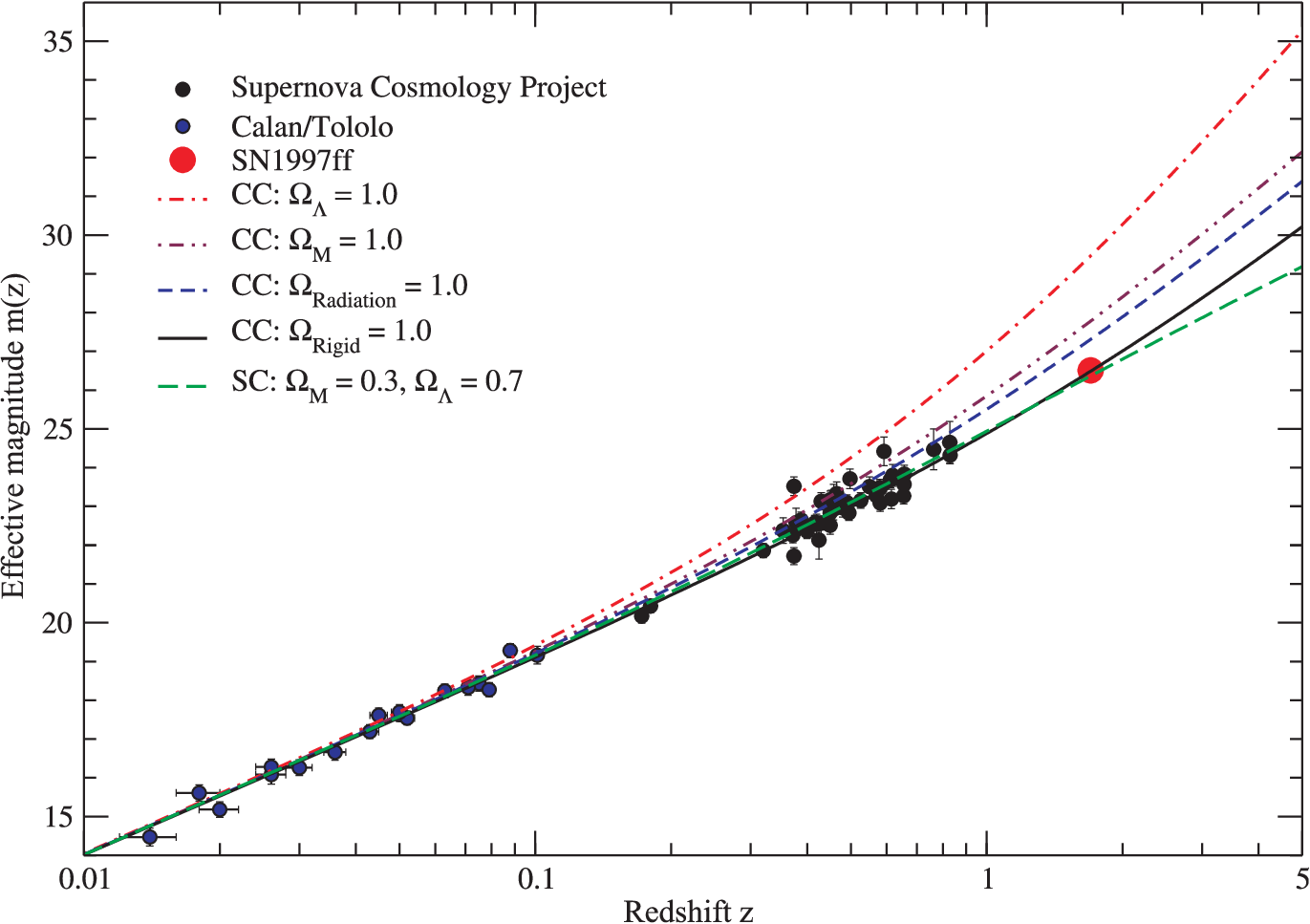}
\caption{
The Hubble diagram~\cite{039a}
in cases of the absolute units of standard cosmology (SC) (\ref{apm}) and
the relative ones of
conformal cosmology (CC) (\ref{rpm}).
 The points include  42 high-redshift Type Ia
 supernovae~\protect\cite{snov} and the reported
 farthest supernova SN1997ff~\protect\cite{sn1997ff}. The best
fit to these data  requires a cosmological constant
$\Omega_{\Lambda}=0.7$ $\Omega_{\rm Cold Dark Matter}=0.3$ in the case of SC,
whereas in CC
 these data are consistent with  the dominance of the rigid state (\ref{data3}).
\label{fig1}}
\end{center}
\end{figure}

In the standard cosmology, an absolute measured distance  is
defined as the product of the scale factor and the coordinate
distance $X^i=a(t)x^i$. This product can be treated as a conformal
transformation that leads to the theory  with constant mass.
%This
%theory is mathematically equivalent to the theory with a relative
%interval and variable mass as all solutions of the second theory
%can be received from those of the first theory by the conformal
%transformation. However, the mathematical equivalence does not
%mean physical equivalence of the absolute and relative units.

When we assert that the cosmic factor was equal to the square root
of ``time'' in the epoch of the primordial nucleosynthesis, the
question appears: what time does an observer measure by his watch,
and what time is identified with the time of chemical evolution?

If this time is absolute, the square root of ``time'' means that
the universe in the epoch of chemical evolution was completed by
radiation when density of pressure is equal to one third of the
energy density.

If this time is conformal, the square root of ``time''
\be\label{data3} a(t)=
\widetilde{a}(\eta)=\sqrt{1+2H_0(\eta-\eta_0)}=\sqrt{1- 2H_0r_c }
\ee means that the universe in the epoch of chemical evolution was
completed by Quintessence when density of pressure is equal to the
energy density.

If we identify  time of the evolution with conformal time and
 substitute the law of  nucleosynthesis (\ref{data3})
 into the Hubble diagram of dependence of
 redshift on distances to Supernovae,
 we can reveal \cite{039} that this law corresponds to the black line in a Fig. 1
 that is in agreement with all data on the Supernova luminocity--distance
--- redshift relation \cite{snov,sn1997ff}.

As it was shown in~\cite{039}, in the case of the {\it relative}
Parisian meter~(\ref{rpm}), both the epoch of chemical evolution
and the recent experimental data for distant supernovae
\cite{snov,sn1997ff} are described by the square root dependence
of the cosmic factor on ``time''. This evolution results from the
dynamics of a homogeneous scalar  field which we call the scalar
Quintessence (SQ). This massless field with purely kinetic
contribution to the energy density in the universe leads to a
rigid equation of state (where pressure is equal to energy) and
gives a satisfactory description of the supernova data.

Other consequence of the relative standard of measurement is the
redshift independence of the cosmic microwave background
temperature \cite{ppgc,114}. This is at the first glance in a
striking contradiction with the observation \cite{sria} of $6.0
~{\rm K} < T_{\rm CMBR}(z=2.3371) < 14~{\rm K}$. However, the
relative population of different energy levels $E_i$ from which
the temperature has been inferred in this experiment follows
basically the Boltzmann statistics with  the same z-dependence of
the Boltzmann factors for both the absolute standards and relative
one  \cite{039}.
 Therefore, the experimental finding can equally well
be interpreted as a measurement of the z-dependence of energy levels (masses)
at constant temperature.

Thus, one more argument in favor of the relative units is
the sharp simplification of the scenario of the evolution of the universe.
Astrophysical data in the relative units can be described by
a single epoch with the dominance of Quintessence, while
the same data in the absolute units require the scenario with
three different epoches (inflation, radiation, and inflation with the dark matter).

All these arguments give a reason of the recalculation of  all
astrophysical data in terms of the relative measurement standard
(conformal time, conformal density, constant
temperature, running mass and others). This recalculation was fulfilled in
\cite{039,plb,039a,ppgc,114,02,295}.

One of the major arguments in favor of the relative measurement standards is
 the symmetry of the theory.
The astrophysical data in the relative units testify to
the hidden conformal symmetry of the Einstein
 general relativity  and Standard Model \cite{ppgc,02,295}.

\section{Conformal symmetry of the world}

Any physical theory beginning with  Newton
at the highest level consists of two parts:
I) differential equations of motion and
II) the initial data which Laplace  still required for unambiguous
solutions of the Newton equations and which  are measured
 by a set of physical instruments identified with  a {\it frame of reference}.

The equations of motion are considered as a {\it kingdom of laws} of  nature;
and the initial data, as a {\it kingdom of freedom}.
In accelerator high-energy physics the experimenters set
the geometry of instruments and initial states of an investigated physical object.
The initial data of the universe are probably set by  Lord-God,
but the essence of theoretical statement of the task remains the same,
and, practically
does not differ from a school task (\ref{np}) about a train moving
in the one-dimensional space with the coordinate $X(\eta) =a^2(\eta)$
with constant speed $V_I=H_0$
from St.-Petersburg $X(0)=X_I=a_I^2$ to Moscow $X(\eta_0)=X_0=1$.

 To find the time dependence of the coordinate of the train
 \be\label{np}
 X(\eta)=X_I+V_I\eta,
 \ee
 it is necessary to solve the Newton equation.
 This equation does not depend on the
initial data (i.e., on the {\it kingdom of  freedom} of passengers
of this train who chose  St.-Petersburg as the initial position
$X_I$
 and the speed of the train $V_I $), but the final
result of the solution of this task - Moscow - is a consequence of
both the {\it kingdoms}: the will of the passengers and  the laws
of Nature. It is important that the Newton equations  do not
depend on the initial data  of the variable $X$.

Independence of the laws of nature on the initial data is called
 the {\it  symmetry of the theory with respect to
 transformations changing the frame of reference},
i.e., rearranging the initial position and speed.

Historically,  frame symmetries appeared as
the Galilean group of transformations rearranging positions and velocities
of the initial data of particles in the Newton mechanics.
The frame symmetry of the modern unified theory
is the Poincare group of transformations
rearranging the initial data of relativistic fields.
 The Poincare  group was recovered by Lorentz and Poincare from the
Maxwell equations. All field theories  of the 20th century were constructed
 by analogy with the  Maxwell electrodynamics ~\cite{p2}.
  In particular, the field nature of  light in electrodynamics and its relativistic
  symmetry were an example for Einstein to formulate his gravitation
  theory. However, an analogy with the Maxwell electrodynamics  was incomplete.

The collection of Faraday's experimental results in the form of
Maxwell's equations testifies to that these  equations are
invariant with respect to conformal transformations\footnote{The
conformal group was discovered by M\"obius in the 19th
century~\cite{w49}, and it was extracted from electrodynamics by
Bateman and Cuningham in 1909~\cite{bat}. The conformal
transformations keep invariant the angle between two vectors in
space-time and include the scale transformations.}.

If to trust that symmetry of the theory of the universe coincides
with symmetry of the theory of light, GR and SM in absolute and
relative units arise as result of the certain choice of gauge in
the conformal--invariant theory of  scalar field called {\it
dilaton},
  \be\label{GR:CI}%
S_{\rm tot}=\int
d^4x\left\{|e|w^2\left[\partial_{\mu}Q\partial^{\mu}Q-\frac{R(e)}{6}\right]
+ w\partial_\mu\left(|e|
\partial^{\mu} w\right)\right\} +  {S_{SM}}[y_hw|f,e],
\ee%
where  the GR action is replaced by the Penrose---Chernikov---Tagirov
action \cite{pct} for the scalar field $w$ called dilaton
 \be\label{GR:CI11}%
S_{\mbox{\tiny PCT}}=\int
d^4x\left\{-|e|\frac{w^2}{6}R(e)+ w\partial_\mu\left( |e|
g^{\mu\nu}\partial_\nu w\right) \right\},
\ee%
 and the Higgs mass in the SM action
  $(M_{\mbox{\tiny Higgs}}=y_h\vh_0)$
  is replaced by this dilaton $(y_h w)$.
 In such a theory all masses are scaled by dilaton  $w$.
Set of fields $F=(Q,f,e)$ including scalar field $Q$ and SM fields
$f$ is given in the space with an interval \be\label{ut1}
ds^2=g_{\mu\nu}dx^{\mu}dx^{\nu}\equiv \omega^2_{\underline 0}-
\omega^2_{\underline 1}-\omega^2_{\underline 2}-
\omega^2_{\underline 3} \ee where
$%\be\label{ut2}
\omega_{\underline \lambda}=e_{\underline \lambda \mu}dx^\mu
$%\ee
are relativistic covariant differential forms with the Fock tetrads
 $e_{\underline \lambda \mu}$.

Action (\ref{GR:CI}) is invariant with respect to
the scale transformations:%
\bea%
g_{\mu\nu} & \to & \widetilde{g}_{\mu\nu}=g_{\mu\nu}\Omega^2  ,\label{ct:g}\\%
w & \to & \widetilde{w}=w{\Omega}^{-1},\\%
{}^{(n)}F & \to &
{}^{(n)}{\widetilde{F}}={}^{(n)}F{\Omega}^{n}\label{ct:f}.
\eea%
The Quintessence $Q$ is treated as the angle of mixing of two
scalar fields $X_0=w\cosh Q$ and $X_1=w\sinh Q$ given in
 the dilaton two-dimensional space with the signature ($+, -$).
 Two dilatons  $X_0,X_1$ and four Higgs field
 $|\Phi|^2=X_2^2+X^2_3+X_4^2+X_5^2$  before the spontaneous
 symmetry breaking describe a  four-dimensional relativistic brane in
 six-dimensional external space with the metrics:
 $$ G_b^{AB}= {\rm diag}(+1,-1,-1,-1,-1,-1).
 $$
 The action of this brane
  \be\label{GR:CB}%
S_{\mbox{\tiny BRANE}}= -\int
d^4x\left\{\sum\limits_{A,B=0}^{5}G^{AB}\left[\sqrt{-g}X_AX_B\frac{R}{6}-
X_A\partial_\mu\left( \sqrt{-g} g^{\mu\nu}\partial_\nu
X_B\right)\right] \right\}
\ee%
belongs to the  class of actions of a relativistic string
  \be\label{GR:CS}%
S_{\mbox{\tiny STRING}}=-\gamma\int
d^2x\sqrt{-g}\sum\limits_{A,B=0}^{D-1}G_s^{AB}\left[
g^{\mu\nu}\partial_\mu X_A
\partial_\nu X_B\right],
\ee%
where $G_s^{AB}=(+1,-1,-1,-1)$ and a relativistic particles
 \be
 \label{SR}
 S_{\rm SR}=-\frac{m}{2}\int\limits_{x^0_1}^{x^0_2}dx^0
\left[\frac{1}{e}G_s^{AB} \frac{d X_A}{dx^0} \frac{d
X_B}{dx^0}+e\right].
 \ee
 There is the unified method of description of energetics of all three
 relativistic theories \cite{Schweber,Logunov}.
 This method is based on the fact
 that groups of transformations of all relativistic actions (\ref{GR:CI}),
  (\ref{GR:CB}),
(\ref{GR:CS}), and (\ref{SR})
 include reparametrizations
of the coordinate evolution parameter $x^0$
$$
x^0 \to \tilde{x}^0=\tilde{x}^0(x^0).
$$
This means that $x^0$ is not observable and the role of measurable
evolution parameter is played by one of the dynamic variables. Its canonical momentum plays the role of energy of a relativistic
system.

\section{The energy of relativistic systems}
\begin{figure}[t]
\vspace{-1cm}
 \begin{center}
\includegraphics[width=0.6\textwidth,clip]{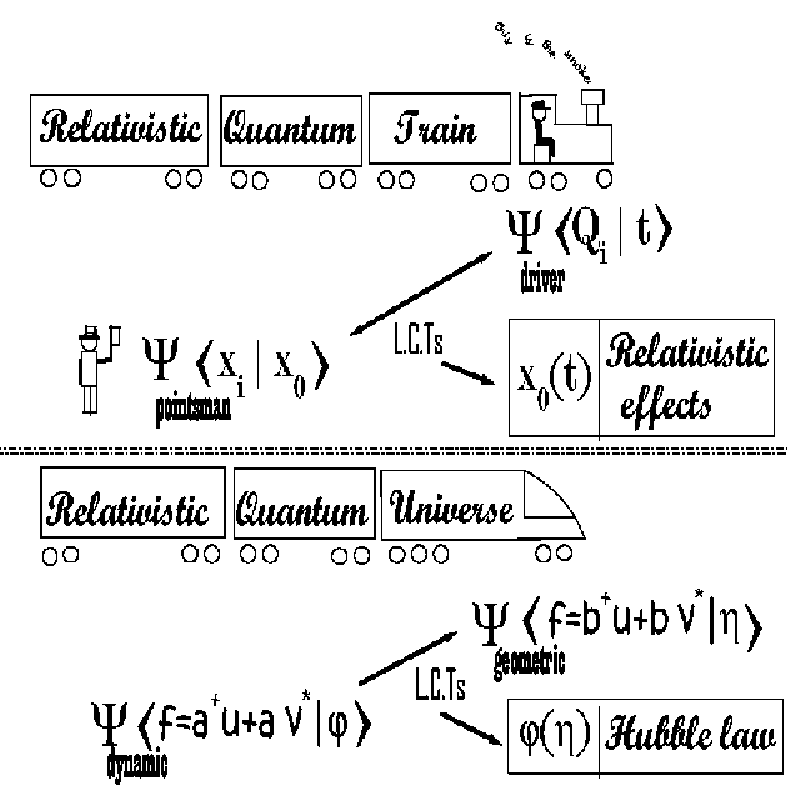}
\caption{{\small At the top of Fig. a relativistic train is
depicted with an unstable particle. The lifetime of this particle
is measured by two Einstein observers, by a pointsman and a
driver, who  communicate to each other their measurement outcomes
on the phone.
  Each of the observers has his  world space of events, his time
 (pointsman - variable $X^0$, and driver - geometrical interval $t$)
 and  his notebook as a wave function of a particle as an amplitudes of
 probability to find a particle at an arbitrary point
 of the world space, if at the initial moment its initial data are given.
 At the bottom of the figure there is an image of
 the universe where each observer has two
 sets of measurable quantities corresponding to two observers of the particle.
 To the pointsman there corresponds a field set of measurable quantities
 (mass $\vh $ and density of a number of particles $n= a^+_qa_q $
 with a set of quantum numbers $q $), and the driver --
  geometrical set of measurable quantities (time interval
  $\eta$  and initial data of the density of the Bogoliubov quasiparticles
 $b_q, b^+_q$)~\cite{ppgc}.
} \label{fig2}}
\end{center}
\end{figure}

In  contrast to  the classical mechanics, in the relativistic
mechanics,  for a complete description  of a relativistic particle
one needs two observers. In  Fig. \ref{fig2} they are depicted in
the role of a pointsman and a driver. A pointsman meters by his
watch the time as {\it a variable} $X_0$ in the world
 Minkowskian space $X_0,X_1$ of all events,
 and a driver meters by his watch the time
as {\it a geometrical interval} $dt=edx^0$ on a world line of
events. Only both sets of measurements
 restore  a pathway of a particle  in the world space $X_0,X_1$ as
 a solution of equations of motion in special relativity
 (\ref{SR}):
$$%\be\label{0np}
X_0(t)=X_{0I}+\frac{P_0}{m}t,~~~~~~~~~~~
%\ee
%\be\label{1np}
X_1(t)=X_{1I}+\frac{P_1}{m}t,
$$%\ee
where the momenta of a particle $P_0,P_1$ are linked by the
mass-shell equation $\delta S_{\rm SR}/\delta e=0$ \be\label{2np}
(P_0)^2-(P_1)^2=m^2. \ee Here the momentum \be
\label{P0:SR:def}P_0={\partial L_{\mbox{\tiny SR}}\over
\partial  \left( dX_0/dx^0\right)}=-m{dX_0\over edx^0}=-m{dX_0\over dt}\ee is treated as an energy of a
particle. Its initial coordinate $X_{0I}$ is treated as the point
of its creation or annihilation in the wave function of a particle
\be\label{f} f(X_0,X_1)=\sum\limits_{q}\left[a^+_q\Psi(X_0\geq
X_{0I})+ a_q\Psi(X_0\leq X_{0I})\right], \ee
 where the coefficients $a^+,a$
are treated as operators of creation, if a particle goes forward,
and of annihilation, if a particle goes backward. This causal
quantization excludes the negative value of the energy $P_0=-E$ to
make stable a quantum state of a particle \cite{Schweber,Logunov}.

   The set of measurable quantities and the wave function of a
   relativistic particle for a driver can be obtained by a transformation
   of the time as the {\it variable} $X_0$ into the time as the
   {\it geometrical interval} $t$ \cite{pp,bpp}.
Such  transformation was firstly proposed in the theory of
differential equations by Levi-Civita as back as  1906 \cite{lc}.
From the point of view of Newtonian physics, the complete
description  of any relativistic object  is possible
 by two  {\it realizations} of this object. For a particle  one of
such {\it realizations} is Minkowskian space, where the  evolution
parameter is the {\it dynamic variable} $X_0 $, and the second is
the geometrical {\it realization}
 where the  evolution parameter is the {\it  time interval} $t$.
  The relationship between these {\it realizations}
is treated as a new, in principle, element of the scientific
explanation of the pure relativistic effects.

The similar choice of the evolution parameter for a relativistic
string \be\label{B-Ch} X_0(x^0,x^1)=X_0(x^0)\ee was firstly
considered by Barbashov and Chernikov \cite{bc}. This gauge
removes excitations of string with negative norm. It was proved
\cite{bpp} that the string theory in this gauge coincides with the
Born---Infeld theory that strongly differs from the Virasoro
algebra~\cite{bn}. Reiman and Faddeev \cite{rf} reproduced and
generalized this result in 1975. In the paper \cite{bpp} the
energetics of a relativistic string was considered on the basis of
the definition of the energy as the canonical momentum of the
evolution parameter $X_0(x^0)$ \be
\label{P0:strig:def}P_0={\partial L_{\mbox{\tiny STRING}}\over
\partial  \left( dX_0/dx^0\right)}.\ee

The description of the universe with the finite volume and lifetime
in the conformal theory
(\ref{GR:CI})  is carried out
in a frame of reference defined by embedding of
three-dimensional hypersurface into
 the four-dimensional Riemannian space-time
\be\label{fr2}
\omega_{\underline{\lambda}}=\left(\omega_{\underline{0}},\omega_{\underline{1}},
\omega_{\underline{2}},\omega_{\underline{3}}\right),~~~~~~~~
\omega_{\underline{0}}=Ndx^0,~~~~~~~~~~~ \omega_{\underline{i}}=
e_{\underline{i}j}(dx^j+N^jdx^0) \ee for an observer at
rest\footnote{An observer moving in the direction of the axis 1
 has: $
 \omega_{\underline{\lambda}}=$ $\Big{(}\omega^v_{\underline{0}}=
\frac{\omega_{\underline{0}}+v\omega_{\underline{1}}}{\sqrt{1-v^2}},
 \omega^v_{\underline{1}}=\frac{\omega_{\underline{1}}+
v\omega_{\underline{0}}}{\sqrt{1-v^2}},
\omega_{\underline{2}},\omega_{\underline{3}}\Big{)}. $  }.

Following Barbashov and Chernikov (\ref{B-Ch}) we can choose an
evolution parameter as homogeneous dilaton with the constant
volume \be\label{ccr} w_{(r)}(x^0,x^i)=\vh(x^0),~~~~~~~~~~~~~~~
V_{(r)}=\int_{}d^3x|\bar{e}_{(r)}|={\rm constant}. \ee
  This gauge excludes all modes of the dilaton with a negative norm
  except of the homogeneous one $\vh(x^0)$ that becomes the
  evolution parameter.

  In the case of the relative units  the conformal invariant action
(\ref{GR:CI}) takes the form
  \be\label{GR:CI1}%
S_{\rm tot}=\int
d^4x\left\{|e_{(r)}|\vh^2\left[\partial_{\mu}Q\partial^{\mu}Q-
\frac{R(e_{(r)})}{6}\right] +
\vh\partial_0\left(\frac{|\bar{e}_{(r)}|}{N}
\partial_{0} \vh\right)\right\} +  {S_{SM}}[y_h\vh|f_{(r)},e_{(r)}].
\ee%

The  theory (\ref{GR:CI}) with the field evolution parameter
$\vh(x^0)$  gives a physical explanation of a problem of
{\bf  horizon} by simultaneous variations of masses of all particles in
the whole three-dimensional hypersurface, as a consequence of the
symmetry of the theory
  with respect to reparametrizations of the coordinate parameter $x^0$
 in the ADM metrics \cite{vlad}.

The theory (\ref{GR:CI1}) has the unambiguous and clear definition
of the {\bf localizable Hamiltonian} of the evolution  as the
canonical momentum of the field evolution parameter $\vh(x^0)$
like (\ref{P0:SR:def}) and (\ref{P0:strig:def})
  \be\label{GR:CI2}%
P_\vh = \frac{\partial L_{\rm tot}}{\partial (\partial_0\vh)} =
-2\int
\omega_{{(r)}\underline{1}}\wedge\omega_{{(r)}\underline{2}}\wedge
\omega_{{(r)}\underline{3}}\frac{d\vh}{\omega_{{(r)}\underline{0}}}
=-2\partial_0 \vh(x^0) \int d^3x \frac{|\bar{e}_{(r)}|}{N}\equiv
-2 V_{(r)}
\frac{d\vh}{d\eta}, %\equiv -2 V_c\vh',
\ee where the Lagrangian is given by the action (\ref{GR:CI1})
$S_{\rm tot}=\int dx^0 L_{\rm tot}$ and $d\eta
=N_0(x^0)dx^0$ is an invariant time interval for для the averaged
lapse function $N^{-1}_0(x^0)=\int d^3x
|\bar{e}_{(r)}|N^{-1}/V_{(r)}$
 \cite{114,pp,bpp}.

In the theory (\ref {GR:CI1}), {\bf homogeneity of the universe}
 is explained by an average of a precise
 equation over the volume, instead of ``inflation''. In particular,
in the theory (\ref {GR:CI1}) the equation of evolution of the
universe
  \be\label{GR:CI4}
\vh'^2=\rho \ee
 where $\rho = \int d^3x |e|[T_0^0-\vh^2(R^0_0-R/2)]$,
 appears as an average of precise equation of the lapse function $N$
  \be\label{GR:CI3}%
 \frac{\delta S_{\rm tot}}{\delta N}=0
\ee over the spatial volume. The solution (\ref{GR:CI4}) is an
analogy of the Friedmann relationship $ \eta(\vh_0,\vh_I)$ $= \pm
\int_{\vh_I}^{\vh_0} {d\vh}/{\sqrt{\rho}} $ in the precise theory
between the time interval and the cosmic scale factor.

The equation (\ref{GR:CI4}) in terms of canonical momentum
(\ref{GR:CI2}) takes the form of the global energy constraint of
the type of the mass-shell equation (\ref{2np})
\be\label{con2}
{P_{\vh}^2\over 4V_{(r)}}=V_{(r)}\rho\equiv V_{(r)}\rho_{\rm
hom}+H_{\rm field}
\ee
where $H_{\rm field}$ is the contribution of
local field excitations and $\rho_{\rm hom}$ is the one of global
homogeneous excitations. In particular, if the homogeneous scalar
Quintessence $Q$ dominates,
  neglecting  all remaining fields,
 we shall receive a homogeneous action
\bea\label{cuf} S_Q&=&V_{(r)} \int \frac{dx^0}{N_0}\left[
\vh^2\left(\frac{dQ}{dx^0}\right)^2\!\!-\left(\frac{d\vh}{dx^0}\right)^2
\right]\\ \nonumber &=& \int \left\{
P_Q\frac{dQ}{dx^0}-P_{\vh}\frac{d\vh}{dx^0}+
{N_0}{V_{(r)}}\left[\left(\frac{P_\vh}{2V_{(r)}}\right)^2\!\!-
\left(\frac{P_Q}{2\vh V_{(r)}}\right)^2\right]\right\} \eea which
in fact is similar to the action of a relativistic particle
(\ref{SR}).

 Cosmology is defined as such homogeneous approximation of GR
 of the type of (\ref{cuf}) that inherits its symmetry
 with respect to reparametrization of the coordinate evolution parameter
 $x^0 \to \widetilde{x}^0=\widetilde{x}^0(x^0)$.
 Reparametrization-invariant cosmological models were firstly
 considered by DeWitt, Wheeler, and Misner
 \cite{dw,M}, do not differ from the special relativity.
  There is a direct correspondence between the Minkowskian space-time
  and the field space of the variables of the model
   (\ref{cuf}) with equations
\be\label{cuf1} P^2_\vh- P^2_Q/\vh^2 =0,~~~~~~~P_\vh=2V_{(r)}
\vh'=\pm P_Q/\vh,~~ ~~~P_Q=2V_{(r)}\vh^2 Q' \ee and their
solutions
 $$P_Q=2V_{(r)}H_I\vh_I^2=\mbox{\rm constant},~~~~~~~~~~~\vh^2=\vh_I^2(1+2H_I\eta),
 $$
 with the dilaton $\vh$ as the evolution parameter.
 Quantum theory, in particular, the Wheeler-DeWitt equation
 \be\label{cufq}
\left[\hat P^2_\vh- \hat P^2_Q/\vh^2\right]\Psi =0 \ee
 appears as a direct analogy of the Klein-Gordon equation
 in relativistic quantum field theory.
The solution of the Wheeler-DeWitt equation \be\label{field1}
\Psi=e^{\{ iP_Q\ln(\vh/\vh_I)\}}\left[A_{{\rm P}_\vh \geq 0}^+
e^{\{ iP_Q(Q-Q_I)]\}}\theta(\vh-\vh_I)+ A_{{\rm P}_\vh \leq 0}^-
e^{\{- iP_Q(Q-Q_I)\}}\theta(\vh_I-\vh)\right]. \ee and its
interpretation do not differ from a similar solution of the
Klein-Gordon equation for a quantum relativistic particle.
 To remove  negative energies and to construct
 a stable quantum system in  relativistic quantum theory
 the causal quantizing of fields is postulated.
 According to this quantizing, the wave with positive
 energy goes forward; and negative energy, backward.
 The same treatment of the coefficient $A^+$
as operator of creation of the universe, and $A^-$, as operator of
annihilation of the universe, solves the problem of {\bf the
cosmic singularity},
 as a wave function of the universe
 with positive  energy  does not contain a point of
 singularity; the singularity is contained in a wave function
 with negative  energy which is treated as a probability amplitude
 of annihilation of anti-universe.

Thus, the conformal unified theory (\ref{GR:CI1}) in the concrete
frame of reference with cosmic initial data gives the possible
solutions of the  problems of modern cosmology.
 At least, these solutions should be considered
on  equal  footing with the old scheme conserving Newtonian
absolute such as the absolute Parisian meter, or the absolute
Planck mass when we fix the gauge of the constant dilaton with the running
 volume
\be\label{cca} w_{(a)}(x^0,x^i)=\vh_0={\rm constant},~~~ ~~V_{(a)}(t)=\int
\omega_{{(a)}\underline{1}}\wedge
\omega_{{(a)}\underline{2}}\wedge\omega_{{(a)}\underline{3}}. \ee This gauge
appears from (\ref{ccr}) by transformation ${}^{(n)}F_{(r)}={}^{(n)}F_{(a)}
(\vh/\vh_0)^{-n}$ that converts the variable $\vh$ with the initial cosmic data
$\vh(\eta=0)=\vh_I$, $H(\eta=0)=H_I$  into its present-day value
$\vh(\eta=\eta_0)=\vh_0$. This transformation creates in equations of motion
the absolute parameter of the Planck mass and the problem of Planck era. The
theory in the gauge (\ref{cca}) loses solutions of problems of cosmic initial
data, horizon, time and energy, homogeneity, singularity,
 and quantum wave function of the universe discussed before. These problems
are solved by the inflationary model \cite{linde}.

\section{``Creation'' of the universe and time}

\begin{figure}[t]
\vspace{1cm}
 \begin{center}
\includegraphics[width=0.55\textwidth,clip]{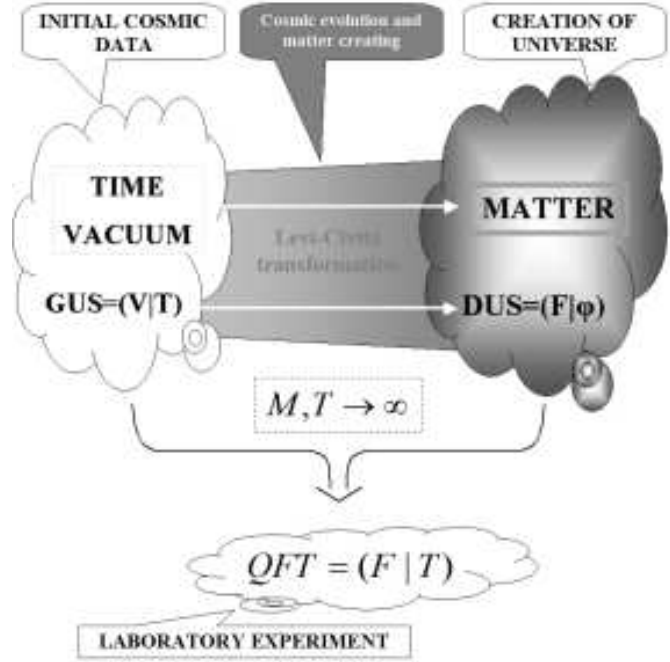}
\caption{{\small This figure, taken from ~\cite{pp}, shows the quantum creation of the universe in the field
reality out of the time
$\eta=T$ (right) and in the geometric reality without matter (left).
The Hubble law $\vh(\eta)$, and creation of matter from vacuum with
 particle density $n(\eta)$ ($n(0)=0$) are
described as pure relativistic effects by the geometrization of the energy constraint.
Only in the limit of tremendous energy of the Quintessence there appears
a possibility of describing the universe as the Newton-like system $F(\eta)$.
}
\label{fig3}}
\end{center}
\end{figure}

The mathematical structure of general relativity and
Standard Model in the relative units (with the evolution parameter $\vh$
and an energy of the universe defined as a value of
the canonical momentum $P_\vh$ of this evolution parameter)
allow us to use the analogy with the relativistic particle  (\ref{f})
to construct a wave function of the relativistic universe in the world field space
  for positive and negative energy   with the initial data $\vh=\vh_I$.

This wave function of the universe
\be\label{field}
\Psi_{\rm field}[\vh,\vh_I|F,F_I]=
A^+_E \Psi_{\rm universe}[\vh \geq \vh_I|F,F_I]+
A^-_E \Psi_{\rm anti-universe}[\vh \leq \vh_I|F,F_I]
\ee
describes the greatest events --
the creation of the universe with  positive energy flying forward
 $\vh \geq \vh_I$ in the field space: the cosmic singularity $\vh = \vh_I$
is in wave function  the universe with  negative  energy
flying backward $\vh \leq \vh_I$ to the point of the singularity.

To make this creation of the universe stable, one
 constructs  the wave function of the quantum universe in the field realization
 excluding the negative value of the energy $P_\vh=-E$ from the wave function.
To do so, we need to treat the creation of the universe with
negative energy as annihilation of the anti-universe with positive
energy. This construction is known in quantum field theory as
causal quantization with the operators of
 creation $A^+$ and annihilation $A^-$ of the
 universe~\cite{Schweber,Logunov}.
Consequences of the causal quantization  (\ref{field}) are the positive arrow
 of the geometric time and its beginning $\eta\geq 0$ \cite{bpp,pp} .

The wave function of the universe in the geometric realization
\be\label{geom} \Psi_{\rm geometric}[\eta\geq 0|G] \ee describes
the quantum evolution of the universe in the geometric world space
$[\eta|G]$ with the zero initial data for matter fields.

The universe  as a relativistic object can also be  completely
described
 by two  {\it realizations}:  field and geometric.
 Each of them has its world space of variables
(field  $\vh,F=(e,f,Q)$, or geometric $\eta,G$), its  evolution
parameter (the cosmic scale factor $\vh$ or geometric time
$\eta$), its initial data, and its wave function (the field
$\Psi_F[\vh\geq \vh_I|F,F_I]$, or  geometric $\Psi_G[{\rm
time}_c=\eta \geq 0|G,G_0]$).

Both these {\it realizations} are connected by the Levi-Civita
transformations that convert  the field space with the field
evolution parameter $\vh$ into the geometric world space
 with the time evolution parameter \cite{pp,bpp}.
The  {\it geometrization}  as a rigorous mathematical construction
of  the geometric time $\eta$ includes the transformations of the
initial fields $F=\sum_q(a_q^+\psi_q+a_q\psi_q^*)$ with a set of
quantum  numbers  $q$ into the
  geometric fields $G=\sum_q(b_q^+\underline{\psi}_q+b_q\underline{\psi}_q^*)$
  known as the Bogoliubov transformations \cite{B,ps1}.

The vacuum initial data $\vh_I, Q_I=0$ including a number of
particles $n_I=\sum_q <0|a^+_qa_q|0>=0$ can be treated as field
coordinates of the creation of the universe in its field
realization. Such a creation takes place out of time $\eta$ that
belongs to another realization of the universe in the geometric
space ($\eta$, $G$).

The evolution of the cosmic scale factor  with respect to time
$\vh(\eta)$
  is considered as a pure relativistic effect  that is beyond the scope of
 the Newton-like mechanics.

 At the beginning of universe there were only two {\it global} excitations
 in the form of
 ``superfluid motions'' (according to the therminology by Landau  \cite{L} and
 Bogoluibov \cite {B}): the running Planck mass   $\vh $
 and Quintessence. The momenta  of these motions  are linked  as the momenta of
 a relativistic particle (\ref{2np}).
 All further evolution of the running Planck mass $\vh(\eta)$
    and measurable number of particles $n_{F}(\eta)\not=0$
    in the field space  $[\vh|F]$ is treated as the Levi-Civita
 geometrization of fields in
 the unified theory $F={F}(\eta,G)$ \cite {pp,bpp}. These transformations
  for local particles coincide with the Bogoliubov transformations
  in his microscopic theory of
 superfluid helium \cite{B}: $a_q=c_q (\eta)b_q+s_q(\eta)b_q^+$.
 In our theory these transformations describe
  cosmological creation of a substance from vacuum in the early universe. The number
  of created particles  is defined as the sum of quadrates of the Bogoliubov
   coefficients
  $s_q (\eta)$:  $n_{F}(\eta)=\sum_q|s_q(\eta)|^2$
 where the magnitude $|s_q(\eta)|^2={\cal N}(q,\eta)$ is called the
 distribution function of the numbers of particles.

\section{Creation of matter}

The origin of particles
 is an open question as the isotropic evolution of the
 universe cannot create massless particles~\cite{ps1,grib}.

 Here we list arguments in favor of that
 the cosmological particle creation from vacuum \cite{grib}
 in the conformal--invariant unified theory can describe
  the cosmic energy density budget of observational cosmology.

At the first moment $\eta_I=1/2H_I$
of the lifetime of the universe, the  frame-fixing quantization \cite{hp}
 of W--, Z-- vector bosons in the Standard Model
 shows us an effect of their intensive cosmological creation \cite{grib} from
 the geometric Bogoliubov
 vacuum~\cite{039,ppgc,114}.
The distribution functions of the longitudinal
$ {\cal N} ^ {||}$ and
transverse $ {\cal N} ^ {\bot}$ vector bosons  calculated in~\cite{039,ppgc,114}
for the initial data $H_I=M_I $
are introduced in Fig. 4.

We can speak about the cosmological creation of a pair
of  massive particles in the universe,  when the particle mass
$ M_v(\eta=0)=M_{I}$  is larger than the initial Hubble parameter
$M_{I} \geq H_I $.

The distribution functions of the longitudinal
$ {\cal N} ^ {||} (x, \tau) $  vector bosons
introduced in Fig. 2 show the large contribution of relativistic momenta.
This means the relativistic dependence of the particle density on the temperature
in the form $n(T)\sim T^3$. These distribution functions show also
that the time of establishment of the density and temperature
is the order of the inverse primordial Hubble parameter. In this case,
one can estimate  the temperature $T$ from the equation
in the kinetic theory \cite{ber} for the time of establishment of the temperature
$$
\eta^{-1}_{relaxation}\sim n(T)\times \sigma \sim H,
$$
where $\sigma \sim 1/M^2$ is the cross-section.

This kinetic equation and values of the initial data $M_I = H_I$
help us to calculate the temperature of relativistic
bosons~\cite{plb,039,ppgc,114}
$$
 T\sim (M_I^2H_I)^{1/3}=(M_0^2H_0)^{1/3}=2.7 K
$$
as a conserved number of cosmic evolution compatible with
the Supernova data \cite{snov,sn1997ff} and the primordial chemical evolution
\cite{three}.
 We can see that
this calculation gives the value surprisingly close
to the observed temperature of the CMB radiation $ T=T_{\rm CMB}= 2.73~{\rm K}$.

\begin{figure}[t]
 \includegraphics[width=0.57\textwidth,height=0.51\textwidth,angle=-90]{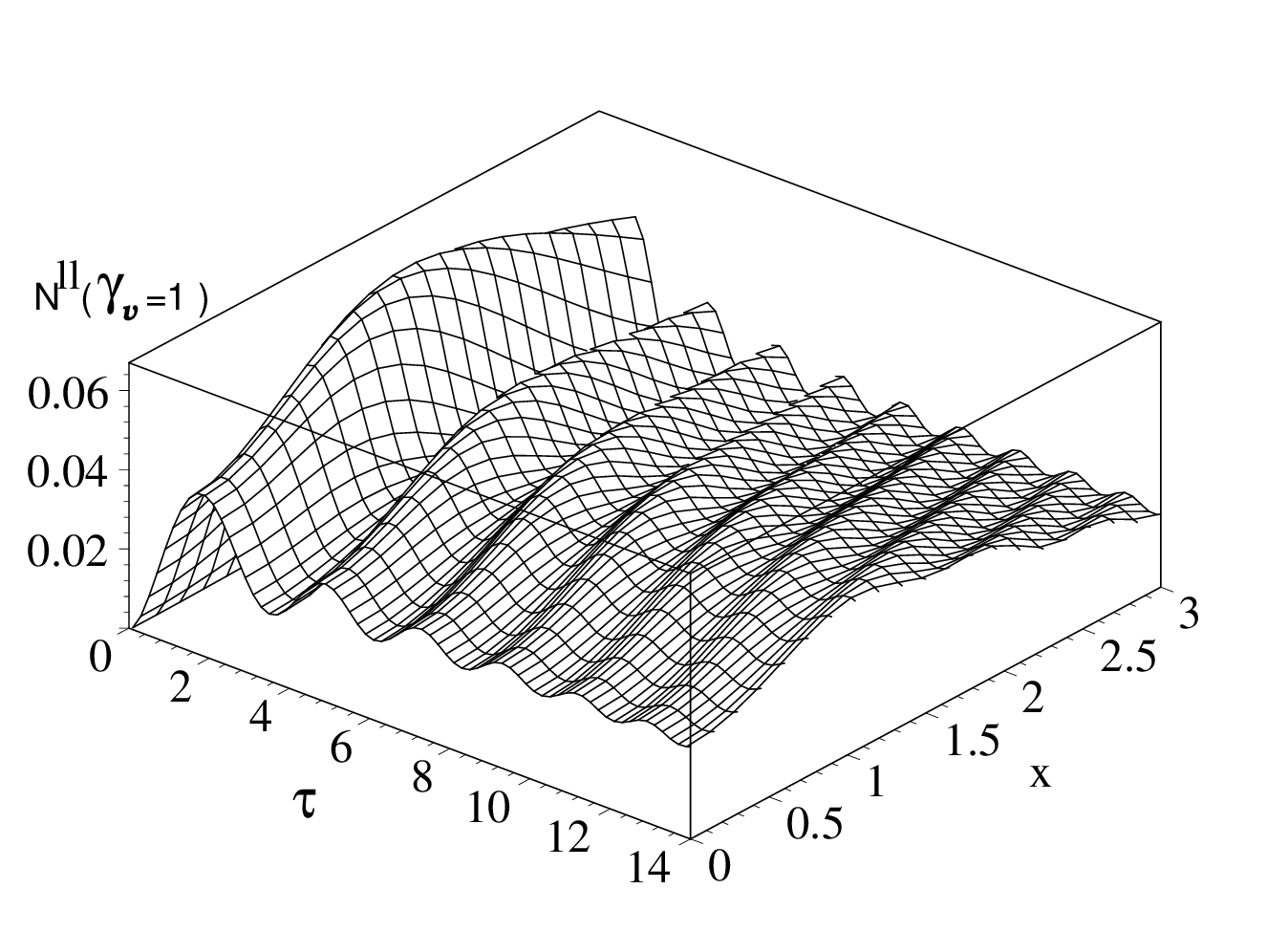}\hspace{-5mm}
 \includegraphics[width=0.57\textwidth,height=0.51\textwidth,angle=-90]{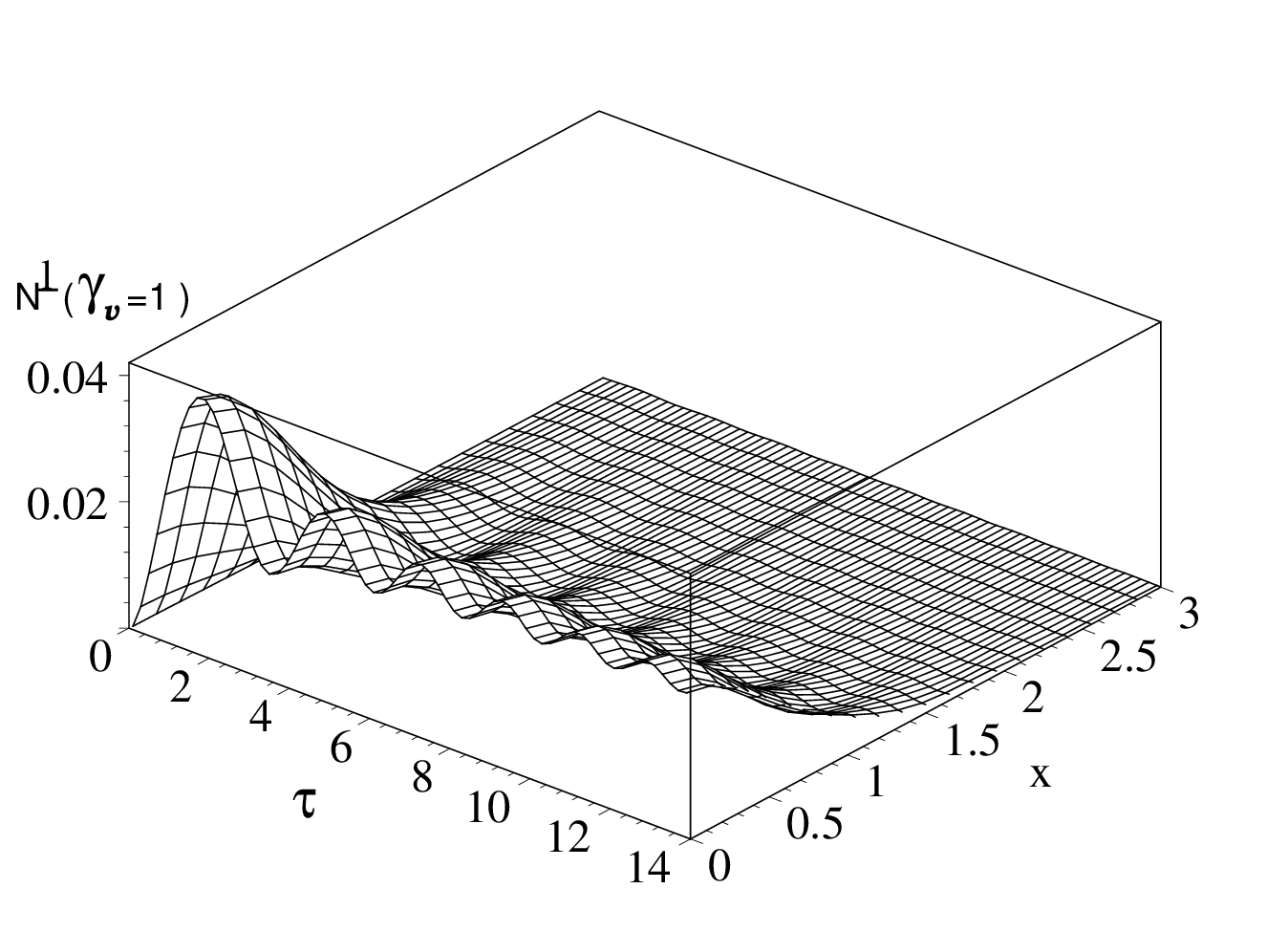}
  \caption{\small The figure taken from \cite{114} shows the dependence of
longitudinal $N^\|$ and transverse $N^\bot $  components of the
distribution function of vector
bosons describing their very fast creation in  units of the primordial lifetime
of the universe $ \tau=2H_I \eta $ \cite{ppgc,114}.
Their momentum distributions in  units of the primordial mass $ x = q/M_I $ show
the large contribution of longitudinal bosons and their
relativistic nature.
}
\end{figure}

A ratio of the density
of the created matter
$ \rho_{\rm v}(\eta_I)\sim T^4$ to the
density of the primordial cosmological motion of the universe
$ \rho_{\rm cr.}(\eta)=H_I^2\vh^2_{I}$ has an
extremely small number
\be\label{data5}
\frac{\rho_{\rm v}(\eta_I)}{\rho_{\rm Q}(\eta_I)}\sim
\frac{M^2_{I}}{\vh_I^2}=\frac{M^2_{W}}{\vh_0^2}
\sim 10^{-34}.
\ee

 On the other hand, it is possible to estimate the
 lifetime of the created bosons in the early universe
 in dimensionless units
$\tau_L= \eta_L/\eta_I $, where $ \eta_I = (2H_I) ^ {-1} $, by
utillizing the equation of state $\vh^2 (\eta_L) =\vh_I^2 (1
+\tau_L) $ and define the lifetime of $W\!$--bosons in the
Standard Model \be \label{lifes} 1+\tau_L= \frac{2H_I\sin^2
\theta_{\rm W}}{\alpha_{\rm QED} M_W(\eta_L)}= \frac{2\sin^2
\theta_{\rm W}}{\alpha_{\rm QED}\sqrt{1+\tau_L}}~, \ee where
$\theta_{\rm W}$ is the Weinberg angle,  $\alpha_{\rm QED}=1/137$.
The solution of equation~(\ref{life}) gives the value for
$M_{vI}\simeq H_I$ \be \label{life} \tau_L+1=
\left(\frac{2\sin^2\theta_{\rm W}}{\alpha_{\rm QED}}\right)^{2/3}
\simeq {16}. \ee The transverse bosons during their lifetime form
the baryon symmetry of the universe
 as a consequence of the ``polarization'' of the
Dirac sea vacuum of left fermions by these bosons,
 according to the selection rules of the
Standard Model~\cite{ufn} with left current interaction in SM
$j_{L\mu}^{(i)}=\bar \psi^{(i)}_L\gamma_\mu\psi^{(i)}_L$ for each
left doublet $\psi^{(i)}_L$ marked by an index $(i)$. At a quantum
level, we have an abnormal current \bea \label{ac}
\partial_\mu j_{L\mu}^{(i)}=-\frac{{\rm
Tr}\hat{F}_{\mu\nu}{}^*\!{\hat{F}_{\mu\nu}}}{16\pi^2}, \eea where
in the lowest order of perturbation theory $\hat{F}_{\mu\nu}=
-({\imath g\tau_a}/{2})(\partial_\mu v_\nu^a-\partial_\nu
v_\mu^a)$. The integration of this equation (\ref{ac}) gives the
number of left fermions $N_L=\int d^4x \sqrt{-g}\partial_\mu
j_{L\mu}^{(i)} $ created during the lifetime
$\eta_L=\tau_L\times\eta_I$ of vector bosons \cite{ppgc,114} \bea
\nonumber
 N_L(\eta_L)=
 -\int_0^{\eta_L} d\eta \int \frac{d^3 x}{16\pi^2} \;
 {}_{\rm sq}\langle 0|{\rm Tr}\hat{F}_{\mu\nu}
 {}^*\!{\hat{F}_{\mu\nu}}|0\rangle{}_{\rm sq}=1.2  n_{\gamma}V,
\eea where $n_{\gamma}=({\pi^2 }/{2.402})T^3$ is the density of
number of the CMB photons. The baryon asymmetry appears as a
consequence of three Sakharov conditions: ${\rm
CP}$-nonconservation, evolution of the  universe $H_0\not= 0$ and
the violation of the baryon number~\cite{114} $ {\Delta B}=X_{\rm
CP}({N_L}/{3V})=0.4X_{\rm CP}n_{\gamma}~, $ where $X_{CP}$ is a
factor determined by a superweak interaction of $d$ and $s$-quarks
$(d+s~\rightarrow ~s+d)$ with the CP-violation experimentally
observed in decays of $K$ mesons with  a constant of a weak
coupling
   $X_{\rm CP}\sim  10^{-9}$~\cite{o}.

After the decay of bosons, their temperature is inherited by the
Cosmic Microwave Background  radiation. All the subsequent
evolution of matter with varying masses in the constant Universe
replicates  the well-known scenario of the hot
universe~\cite{three}, as this evolution is determined by the
conformal-invariant ratios of masses and temperature $m/T$. As the
baryon density increases as a mass and the Quintessence density
decreases as an inverse square mass, the present-day value of the
baryon density can be estimated by the relation \be\label{data7}
\Omega_{\rm b}(\eta_0)=\left[\frac{\vh_0}{\vh_L}\right]^3
\frac{\rho_{\rm b}(\eta_{\rm L})}{\rho_{\rm Q}(\eta_{\rm L})}=
\left[\frac{\eta_I}{\eta_L}\right]^{3/2} \sim
\left[\frac{\alpha_{\rm QED}}{\sin^2 \theta_{\rm W}}\right] \sim
0.03, \ee if the baryon asymmetry with the density $ \rho_{\rm
b}(\eta=\eta_{\rm L}) \simeq 10^{-9} 10^{-34}\rho_{\rm
Q}(\eta=\eta_{\rm L}) $ was frozen by the superweak interaction.
This estimation gives the value surprisingly close to the
observational density in  agreement with the observational data.
In the current literature \cite{mar} the cosmological creation of
particles is considered as an origin of the primordial
fluctuations of temperature of CMB \cite{33}. Generally speaking,
all these present and future results can only be treated as a set
of arguments in favor of the considered unified theory.

Thus, we have shown that the conformal-invariant version of general relativity
and Standard Model with
geometrization of constraint and frame-fixing with
the primordial initial data $\vh_I=10^{4} {\rm GeV}$,
$H_I= 2.7~ {\rm K}=10^{29}H_0$
(determined by a free homogeneous motion of
the Scalar Quintessence, i.e., its electric tension)
can describe the following
events:
$$
\begin{array}{rll}
&\eta =0 &{\rm creation~ of~ the ~ ``empty"~universe~from ~``nothing"}\\ [1.5mm]
&\eta \sim 10^{-12}s &{\rm creation~ of~ vector~bosons~from ~``nothing"}\\ [1.5mm]
10^{-12}s <\!\!\!\!&\eta < 10^{-11}\div 10^{-10} s~
~~&{\rm formation~ of~ baryon~asymmetry}\\ [1.5mm]
&\eta \sim 10^{-10}s&{\rm decays~ of~ vector~bosons}\\ [1.5mm]
10^{-10}s <\!\!\!\!&\eta < 10^{11}s&
{\rm primordial~ chemical~ evolution ~of~matter}\\ [1.5mm]
&\eta \sim  10^{11}s&{\rm recombination,~or~separation ~of ~CMB}\\ [1.5mm]
&\eta \sim  10^{15}s&{\rm formation~of~galaxies }\\ [1.5mm]
10^{17}s <\!\!\!\!&\eta&{\rm hep~experiments~and~Supernova ~evolution}.
\end{array}
$$

 The key differences of such a description from
 the inflationary model \cite{linde}
  are the absence of the Planck era and
   cosmological creation of matter from the physical
  vacuum  as a stable state with lowest energy at the moment when
  the size of the horizon of events in the universe coincides
  with the Compton length of
  the vector W-,Z-bosons.

\section{Conclusion}

The relative measurement standard  opens out the well-known truth
that the universe is an
  ordinary physical object with a finite volume
 and finite lifetime\footnote{
 It demonstrates us the dark sky  at night, to what
 Halley, de Ch\'eseaux, and Olbers paid attention \cite{lang}.}.

Results of theoretical description  of the finite universe depend on the
choice of a frame of reference and initial data like
the results of  solution of the Newton equations
  depend on initial positions and initial velocities of a particle.
 Creation of the universe has taken place in a particular frame of reference
 which was remembered by the cosmic microwave radiation.
 We  remind that the ``frame of reference'' is identified with a set of the
 physical instruments for measuring the initial data needed for  unambiguous
   solving differential equations of theoretical physics. These
differential equations are invariant structural relations of the whole
manifold of all measurable quantities with respect to their transformations.
The determination of a group of these transformations is the most
important problem of the modern theoretical physics.

    There are two types of the
 transformation groups of differential equations of the
 gauge theory: {\it frame-transformations} that change initial data,
 i.e., the {\it frame of reference}; and {\it gauge-transformations}
 that do not change initial data and are
 associated with the calibration of physical instruments.

   Derivation of  frame-covariant and gauge-invariant solutions of differential
 equations as well as the construction of frame-covariant and gauge-invariant
 quantization of
 gauge fields were considered as the mainstream of
  development of
 theoretical physics beginning with the work by Dirac \cite{1}
 and ending with the work by Schwinger in the sixties who called
 this quantization {\it fundamental} \cite{sch2}. The strategy of this
 {\it fundamental quantization}  was to construct
 gauge-invariant variables in a definite frame of reference and to prove
 the relativistic invariance of a complete set of results \cite{sch2,pol,mpl}.

The basic method of quantization in gauge field theories, however,
 became the other {\it heuristic quantization}, proposed by
 Feynman~\cite{Feynman1,Feynman2}. Feynman noticed that the scattering
 amplitudes of the elementary
 particles  in  perturbation theory  do not depend on the frame of
 reference and the gauge choice~\cite{Feynman1}.
 The independence of the frame of reference
 was called simply the {\it relativistic invariance},
 and the gauge choice became the formal procedure of choosing
 the gauge non-invariant field variables. It may
 seem that this slight substitution of the meaning of
 the concepts in the method {\it heuristic quantization} completely depreciates
 the goals and tasks of the
 {\it fundamental quantization}. Why should we prove the
 ``relativistic invariance`` of a complete set of results
  at the level of the algebra of the
 Poincare group generators for gauge-invariant observables,
 if the result of calculating of each scattering amplitude is
 relativistic invariant, i.e. does not depend on a frame of reference \cite{fp1}?
 What do we need gauge-invariant observables for,
 if one can use any variables also for solving the
 problems of construction of the unitary perturbation theory
 and proving the renormalizability of the Standard Model \cite{hw}?
 The statement and solution of these important problems
 carried out  within the limits of {\it heuristic quantization}
 resulted in that the latter became one and the only method of solving
 all the problems of the modern field theory.
 The highest achievements of the {\it abstract} formulation
 are the frame-free quantization of string theories
 and M-theory as a candidate for the role of a future
 consistent theory of all interactions  with the
 Planck absolute mass (see, e.g., \cite{duff}).

 At the end of the past century, a dramatic situation arose in physics,
   when a historical path of physics - the path of the
  frame of references,  seems to be absolutely interrupted.
 Remained only  the ``kingdom of laws'' burdened with absolutes independent from a
  frame of reference. The ``kingdom of freedom'' of initial data
 turned out to be enclosed  by {\it heuristic quantizing} and its claims
 for a successful solution of all problems.
  There was a new terminology
  with the distorted definition of relativistic invariance, suitable
  only for the description of the tasks of scattering.

  However, physicists have forgotten that the simplified {\it heuristic quantizing}
  is proved only for amplitudes of scattering of elementary
 particles \cite{f1},
 and its applicability is restricted to only scattering problems
 -- the domain where it first appeared \cite{Feynman1}.
 The {\it fundamental quantization} is more suitable for
 the physics of bound states,
 hadronization and confinement, and
  for  the description of the quantum universe  \cite{pp,mpl}.
 Yet in 1962, Schwinger \cite{sch2} pointed out that
  the frame-free formulations can
  distort the initial gauge theory and lead to a wrong spectrum of
 nonlocal collective excitations.
 Schwinger
 rejected all  frame-free formulations
 of relativistic theories ``{\it as unsuited to the role of providing the
 {\it fundamental} operator quantization}''~\cite{sch2}\footnote{
 Frame-free formulations lead to another spectrum of the nonlocal
 bound states: when we  replace in QED
 the Dirac  sources in the Coulomb gauge of by the Lorentz ones, to
 acquire the independence on any frame of reference and
 initial data, we substitute the perturbation  theory
 of  fundamental quantization with the two singularities
 of photon propagators: at the equal time
 instants
 and on the light  cone, by the
 perturbation  theory in the Lorentz gauge  with only one
 singularity of photon propagators. The latter kind
 of singularities cannot, in principle, describe the Coulomb atoms
 at the equal time instants; these singularities
 describe only the  Wick-Cutkosky bound states
 \cite{Kummer} with the bound state
 spectrum which are impossible to be observed
 in the nature.}.

In 1974, Barbashov and Chernikov \cite{bc}
applied the frame-fixing formulation to the  relativistic string theory
and proved that this theory coincided with the Born-Infeld theory
that strongly differs from the abstract frame-free formulation
of a string with the so-called  Virasoro algebra~\cite{bn}.
Reiman and Faddeev~\cite{rf}
reproduced and generalized this result in 1975 (for details see~\cite{bpp}).

 The relative measurement standard reverts us on a historical path
 of physics, the path of frame of references. This path began
 with relativity by Copernicus, Galilei, and Newton,
   and it was prolonged by Einstein's relativity theories and papers by
  Dirac, Heisenberg, Pauli, Fermi and Schwinger on gauge-invariant
  {\it fundamental quantizing}.
 It is the path of definition of a transformation group
  of all measurable physical quantities, which leaves
 invariant their structural relations called
 the differential equations.
 It is the path where all absolutes of theories become,
  eventually, ordinary initial data.

 The relative units reveal  that  a symmetry group of the whole
  manifold  of measurable physical quantities in the world includes
 conformal symmetry of the Faraday-Maxwell electrodynamics,
 and the field nature of matter should also be supplemented
 by the field nature of space and time.

 The relative units lead us to the ``kingdom of freedom''
 of initial data  including also last dimensional absolute of modern
  quantum field theory  and those initial data of creation
 of the universe, for which an observer does not carry any responsibility,
 as he at this moment existed only as an intention.
 Who has carried out this experiment of creation of
  the universe? Who has determined the initial data of this creation?
 Whose notebook is the wave function of the universe?

%\vspace{1cm}
\section*{Acknowledgment}
Authors are  grateful to Profs. B.M. Barbashov, D. Blaschke, N.A. Chernikov,
P. Flin, J. Lukierski, M.V. Sazhin, and  A.A. Starobinsky  for fruitful discussions.
This work was supported in part by the Bogoliubov-Infeld Programme.

\appendix
\renewcommand{\thesection}{Appendix \Alph{section}:}
\renewcommand{\theequation}{\Alph{section}.\arabic{equation}}

 \section{Penrose--Chernikov--Tagirov
 theory in \\ the Barba\-shov--Chernikov gauge}
\setcounter{equation}{0}

 We consider the dilaton part of the ``brane''  (\ref{GR:CI})
 described by the PCT action (\ref{GR:CI11}) with negative sign
 \be\label{S:PCT:negative}
 S_{\mbox{\tiny PCT}}=\int d^4x\left\{-|e|\frac{w^2}{6}R(e)+
 w\partial_\mu\left( |e| g^{\mu\nu}\partial_\nu w\right) \right\},
 \ee
 The absolute gauge $w_{(a)}=\vh_0$ means the choice of the
 variables by the scale transformation
 \be
 e_{(a)i\underline{a}}=\left({w\over \vh_0}\right)
  e_{\underline{a}i}, ~~~~w_{(a)}=\left({w\over
 \vh_0}\right)^{-1}w=\vh_0 \ee
 that leads to Einstein`s
general relativity
 \be S_
 {\mbox{\tiny
 PCT}}[w=\vh_0|e]=S_{\mbox{\tiny GR}}[\vh_0|e]=-\int d^4 x |e| {
 \vh_0^2 \over 6}R(e)
 \ee
This gauge introduces the absolute
parameter $\vh_0$ into equations of motion together with absolute
Planck era.

The Hamiltonian approach to general relativity
is well known  \cite{dw,M}
 \be\label{GR:1}
 S_{\mbox{\tiny GR}}[w=\vh_0|e]=\int
 dx^0\left\{\left[\int d^3xP_{\underline{a}}^i\partial_0
 e_{\underline{a}i}\right]-H_{\mbox{\tiny GR}}[\vh_0|e]\right\},
\ee
where
\be\label{GR:11}
 H_{\mbox{\tiny GR}}[\vh_0|e]= \int d^3x \left\{N{\cal H} - N^k {\cal P}_k
 +C_0[P_{\underline{a}}^i
 e_{\underline{a}i}]+C_{\underline b}f_{\underline b}(e)\right\}
 \ee
is the standard Hamiltonian with the constraints, gauges, and
  the Lagrangian multipliers $N$, $N^k$, $C_0$, $C_{\underline b}$ discussed
  in \cite{dw,M,fp2}.

The Barbashov--Chernikov gauge \cite{bc,rf,bpp}  $w_{(r)}=\vh(x^0)$
means the choice of the relative variables by the scale
trasformation\be e_{(r)i\underline{a}}=\left({w\over \vh}\right)
e_{\underline{a}i}, ~~~~w_{(r)}=\left({w\over
\vh}\right)^{-1}w=\vh, \ee
that keeps the evolution parameter as a
dynamic variable in correspondence with the invariance of the
theory under reparametrizations of the coordinate evolution
parameter.

PCT theory in the terms of relative variables takes the form
 \be \label{act}
 S_{\mbox{\tiny PCT}}[\vh|e_{(r)}] = S_{\mbox{\tiny GR}}[\vh|e_{(r)}]
 + S_{\mbox{\tiny INTERFERENCE}} +
 S_{\mbox{\tiny UNIVERSE}},
 \ee
 where the first term coincides with the Einstein action
 with the Hamiltonian {GR:11}
\be\label{GR:111}
 H_{\mbox{\tiny GR}}[\vh|e_{(r)}]= V_{(r)}\rho(\vh)
 \ee
in terms of relative fields and {\it dynamic evolution parameter}
$\vh$ instead of the Planck mass $\vh_0$; the second term
 \be \label{interf}
S_{\mbox{\tiny INTERFERENCE}} = - \int\limits_{x^0_1 }^{x^0_2
}dx^0 {\partial_0 (\vh^2) } \int
 \limits_{V_0}d^3x P_{\underline{a}}^i
e_{\underline{a}i}
 \ee
is interference  between the global motion of the universe and the
local field excitations; the third term
 \bea \label{universe}
S_{\mbox{\tiny UNIVERSE}} &=&  - V_{(r)}\int\limits_{x^0_1 }^{x^0_2 }dx^0
\frac{(\partial_0 \vh)^2 }{\bar N_0}=-\int^{x^0_2}_{x^0_1} dx^0
\left[ P_{\vh}\partial_0\vh+N_0{P_{\vh}^2\over 4V_{(r)}}\right]
 \eea
is the action of the global motion of the universe in the field world space,
 $N_0[ e, N]$
 and $V_{(r)}$ are considered as
   functionals given by
 \be\frac{1}{  N_0}=\frac{1}{V_{(r)}}\int d^3x
 \frac{ |{}^{(3)} e_{(r)}|}{ N},
 \qquad V_{(r)}=\int d^3x|{}^{(3)} e_{(r)}|~.
 \ee
 The interference of the global and local variables disappears,
 if we impose the Dirac condition \cite{d1}
 of the minimal embedding of the three-dimensional hypersurface
 into the four-dimensional space-time in the relative space $P_{\underline{a}}^i
 e_{\underline{a}i}=0$.
 The minimal embedding removes  not only the interference
 of the global motion with local excitations, but also
 all local excitations with the negative norm~\cite{fp2}.

The PCT theory in terms of relative variables is the direct field
generalization of SR with two time--like variables (the geometric
interval $d\eta=N_0dx^0$ and {\it dynamic evolution parameter}
$\vh$) and two wave functions. We have one to one correspondence
between SR and PCT theory \cite{ps1,pp}, i.e., their proper times
\be\label{corrt}
         dt=ed\tau ~~~~\Longleftrightarrow~~~~~~~
          d\eta=  N_0dx^0,~~~~~\ee
 \rm their world spaces\be
 \label{corrw}
       ~~~ X_0,~X_i  ~~~~~~   \Longleftrightarrow ~~~~~~~
           \vh, ~ e_{\underline{a}i},~~~~~~~~~~ ~~~~~
 \ee their energies
 \be\label{corre}
 ~~~P_0=\pm\sqrt{P_i^2+m^2}~~~ ~~\Longleftrightarrow~~~~~~~
 ~ P_{\vh}=
 \pm 2 V_{(r)}\sqrt{\rho(\vh)} ,~~~~~~~~~
 \ee
 and their two-time relations in
the differential form
\be  \label{corrdt}
~~ \frac{dX_0}{dt} =\pm\frac{\sqrt{P_i^2+m^2}}{m}~~~ ~~\Longleftrightarrow~~~~~~
  ~\frac{d\vh}{d\eta} = \pm
\sqrt{\rho(\vh)},~~~~~~~~~~~~~~~~~~
\ee
and in the integral forms \be \label{corrit}
  ~~~~~~t(X_0)=\pm\frac{m}{\sqrt{P_i^2+m^2}}X_0  ~~~ ~~\Longleftrightarrow~~~~~~~~
\eta(\vh_0,\vh_I) = \pm
\int\limits_{\vh_I}^{\vh_0}\frac{d\vh}{\sqrt{\rho(\vh)}}~.~~~~ \ee

Recall that in SR eq. (\ref{corrit}) can be treated as the Lorentz
 transformation of the rest frame with time $X_0$ into the comoving one
 with the proper time $s$. Similarly, in GR  the relation
 $\eta(\vh)$ defined by eq. (\ref{corrit}) is treated
  as a canonical transformation \cite{pp}. This  correspondence
  (\ref{corrt})-(\ref{corrit}) allows us to solve the problem of time
  and energy  in GR  like Poincare and Einstein \cite{poi,ein}
   had solved it in SR. They identified the time with one of variables
   in the world space. The similar String/SR correspondence was
   considered in papers \cite{bp,bpp}.

\appendix
\renewcommand{\thesection}{Appendix B:}

\section[~~~~~~~~~~~~~~The cosmic  energetics of Galaxies]{The cosmic
 energetics of Galaxies}\label{ut3-2}
\setcounter{equation}{0}
\renewcommand{\theequation}{B.\arabic{equation}}

In this Appendix the solution of the Kepler problem is
given in the Friedman--Robertson--Walker metric \cite{543}.
Here we show that the cosmic evolution depresses an energy of
particles, urging free particles to be captured in bound states,
and free galaxies, in clusters of galaxies.

The formulation of the Kepler problem in the
 Friedmann---Robertson---Walker (FRW) metrics
 \be\label{gra1} {
 ds}^2={(dt)^2-a^2(t)(dx^i)^2}
 \ee
 proposes a choice of physical variables, coordinates, and units of measurement.
  In particular, the choice of absolute units of the expanding universe
 means that the coordinates $X^i=a(t)x^i$ are measurable.
 In terms of these
 coordinates the interval (\ref{gra1}) becomes
 \be\label{gr3a1} {
 ds}^2={(dt)^2-a^2(t)(dx^i)^2}\equiv
 {(dt)^2-(dX^i-HX^idt)^2},
 \ee
 where $H=\dot a(t)/a(t)$ is the Hubble parameter,
 and $HX^i$  are the Hubble velocities that should be taken into
 account in the energy of matter in the universe.
 The Hubble velocities are contained in  the covariant derivatives in
 the Newton action in the space with the interval (\ref{gr3a1})
 \be\label{cr11a} {
 S_A=\int\limits_{t_I}^{t_0}dt\left[P_i(\dot X^i-HX^i)-
 \frac{P_i^2}{2m_0}+\frac{\alpha}{R}\right], }
 \ee
 where
 $\alpha={M_\odot m_I G}$  is a
  constant of a Newtonian
 interaction of a galaxy with a mass $m_I$ in a gravitational field of mass
  of a cluster of galaxies ${M_\odot}$.

  The last three summands
 in this action (\ref {cr11a}) are identified with
 the Hamiltonian of a ``particle'', a value of
 which on solutions of  equations of motion
 is called an energy of the system. It is easy to see that
   the Hubble velocities in action (\ref {cr11a}) are contained in
  the additional summand in the total energy of the system  (\ref{cr11a})
\be\label{def:H}
 E_{\rm tot}=HRP_R+E_N,
 \ee
 as contrasted to a customary Newtonian energy of the system $E_N$
 $$
 E_N=\frac{P_R^2}{2m_I}+\frac{M^2_I}{2m_IR^2}-\frac{\alpha}{R},
 $$
 for the circular velocity $v_I$, initial radius $R_I$, and orbital momentum
  $M_I=v_IR_Im_I$ in the cylindrical coordinates
 \be
 \label{coord}
 X^1=R\cos\Theta,~~X^2=R\sin\Theta,~~X^3=0.
 \ee
 These components of Hubble's velocity $HX^i$ are not taken into account
 in  papers  \cite{ch,E:1,E:2,pri} analysing the problem
 of Dark Matter on the basis of the Newtonian motion of a particle
 in the gravitational field.

 Let us consider a solution of the Kepler problem
 in the expanding universe $H\not =0$ for the rigid state
 \be\label{hab2}
 a(\eta)^2=a(\eta_I)^2[1+2H_I(\eta-\eta_I)]
 \ee
 which describes the recent Supernova data \cite{snov,sn1997ff} in
 the relative units.

 Knowing the link between
 the Friedmann time $t$ and the conformal one $\eta$, and  the scale factor
 it is possible to find the magnitude $H(t)$
 \be\label{hab1}
 H(t)={H_I\over 1+3H_I (t-t_I)}.
 \ee

 It is worth  reminding that the energy conservation law
  $\dot{E}_N(t)=0$ in the Newton theory in the flat space-time
  gives the link of the initial data $v_I$, $R_I$
 $$
 v_I=\sqrt{\frac{\alpha}{m_IR_I}}\equiv v_N,
 $$
 so that the energy of a particle is always negative for all its initial data
 $$
 {E}_N=-m_I\frac{v_N^2}{2}\leq 0.
 $$

\begin{figure}[ht]
\begin{center}
\includegraphics[width=0.4\textwidth,height=0.1\textheight]{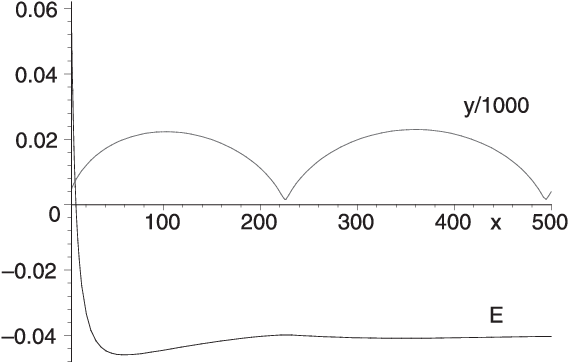}
\end{center}
\caption{\small The  numerical solution of the equation
for the theory (\ref{cr11a}).
 In dimensionless variables $y(x)=R/R_I$ and $x=H_I(t-t_I)$
 with boundary conditions
$y(x=0)=1$ and  $y'(x=0)=0$. The curve at the bottom of the figure
demonstrates the
 evolution of the total energy (\ref{def:H}).}
\label{fig:h-x}\end{figure}

\begin{figure}[htb]
\begin{center}
\includegraphics[width=0.35\textwidth,height=0.1\textheight]{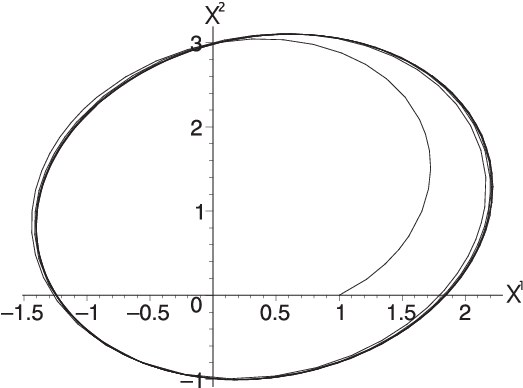}
\end{center}
\caption{\small Pathway of a motion of a particle in the cartesian
coordinates $(X^1,~X^2)$ which  starts with zero value of the
total energy (\ref{def:H}).} \label{fig:y-phi}\end{figure}

 In the considered case of the nonzero Hubble velocity one has that
 the initial total energy
 $$
 {E}_{\rm tot}(t_I)=m_I\frac{(H_IR_I)^2-v_N^2}{2}.
 $$
 becomes positive in the range of large radiuses  $R_I=R(t_I)$ at finite $t$,
 and the link of the initial data at sluggish variation of
 the Hubble parameter takes the form
 $$
 v_I=\sqrt{v^2_N+2(H_IR_I)^2}.
 $$
 It  means, that for large  radiuses
 \be \label{cr}
  R_I\geq R_{\rm cr}=\left(\frac{\alpha}{H_I^2m_I}\right)^{1/3}
 \ee
 a galaxy becomes  free.
 One can see that the critical radial distance (\ref{cr}) is very close to the
 size of Galaxies, and it even coincides with the size of the COMA
 $R_{\rm COMA}\sim R_{\rm cr}\sim 5 \cdot 10^{25}cm$ \cite{ch,E:1,E:2}.
 Thus, just in the region of the expected halos \cite{ch,E:1,E:2,pri}
 we have the cosmic evolution of Galaxies.
% The cosmic evolution can imitate Dark Matter halos
% beyond the validity region of the above Newton approximation, i.e.,
% at $R_I\geq R_{\rm cr}$.

 Let us consider  a case of a ``particle''  with initial data at $t=t_I$ with
  a zero-point energy $E_{\rm tot}(t_I)=0$ given by (\ref{def:H}) in the
  form
 \be\label{def:H1}
 E_{\rm tot}(t)=\frac{m_I[\dot R^2(t)-H^2(t)R^2(t)]}{2}
  +\frac{M^2_I}{2m_IR^2(t)}-\frac{\alpha}{R(t)}
\ee
 in the solution of the  equation of motion
 \be\label{def:H2}
 m_I\ddot R(t)+2{m_IH(t)^2 R(t)}
 -\frac{M^2_I}{m_IR^3(t)}+\frac{\alpha}{R^2(t)}=0.
 \ee

 Solution of the equations for the case of (\ref{hab1})
 gives a remarkable fact:
 a bit later $t\geq 275 t_I $ this particle
  acquires negative energy $E _ {tot} = -0.0405$
(as shown in  Fig. \ref {fig:h-x}) and  also becomes bound

 That is, the cosmic evolution   forms
 the Kepler  bound states such as galaxies and their clusters.
The cosmic evolution depresses energy of fragments, urging free
fragments to capture in bound states, and free galaxies, in
clusters of galaxies.

\end{document}